\documentclass[10pt,a4paper]{article}
\usepackage[dvips]{graphics}

\begin{document}

\hfill\parbox{30mm}{\normalsize DPNU-01-15}

\vspace{5mm}
\begin{center}
{\Large\bf Understanding the penguin amplitude in $B\to\phi K$ decays}

\vspace{5mm}
S. Mishima
\footnote{e-mail: mishima@eken.phys.nagoya-u.ac.jp}

{\it Department of Physics, Nagoya University, Nagoya 464-8602, Japan}
\end{center}

\vspace{5mm}
\begin{abstract}
We calculate branching ratios for pure penguin decay modes, 
$B\to \phi K$ decays using perturbative QCD approach. 
Our results of branching ratios are consistent with the experimental 
 data and larger than those obtained from the naive factorization
 assumption and the QCD-improved factorization approach. 
This is due to a dynamical penguin enhancement in perturbative QCD
 approach.   
\end{abstract}

\begin{center}
{\bf PACS index : 13.25.Hw, 11.10.Hi, 12.38.Bx}
\end{center}

%
%

\section{Introduction}\label{sec:intro}
Recently the branching ratios of $B\to \phi K$ decays have been
measured by the BaBar \cite{BABAR}, BELLE \cite{BELLE} and 
CLEO \cite{CLEO} collaborations.  
There is an interesting problem related a penguin contribution to 
decay amplitudes \cite{PEN}. 
A naive estimate of the loop diagram leads to $P/T \sim \alpha_s/(12\pi) 
\log(m_t^2/m_c^2)\sim O(0.01)$ where $P$ is a penguin amplitude and $T$ 
is a tree amplitude. But experimental data for ${\rm Br}(B \to
K\pi)$ and ${\rm Br}(B \to \pi\pi)$ leads to $P/T\sim O(0.1)$.  
Therefore, there must be a dynamical enhancement of the penguin
amplitude. This problem is studied by Keum, Li and Sanda using
perturbative QCD (PQCD) approach \cite{KLS}. 
$B\to \phi K$ modes are important understanding penguin dynamics,
because these modes are dominated by penguins. 
Here we report our study of $B\to \phi K$ decays using PQCD.

PQCD method for inclusive decays was developed by many authors over many
years, and this formalism has been successful. Recently, PQCD has been
applied to exclusive B meson decays,
$B \to K\pi$ \cite{KLS}, $\pi\pi$ \cite{LUY}, $\pi\rho$, 
$\pi\omega$ \cite{LY}, $KK$ \cite{KK} and $K\eta^{(')}$ \cite{KoSa}.   
PQCD approach is based on the three scale factorization
theorem \cite{CL},~\cite{YL}. 
For example, $B \to K$ transition form factor can be written as
\begin{equation}
F^{BK} \sim \int [dx][db] C_i(t) \Phi_K(x_2,b_2) H(t)\Phi_B(x_1,b_1)
\exp\left[-
\sum_{j=1,2}
\int_{1/b_j}^t\frac{d{\bar\mu}}
{\bar\mu}\gamma_\phi(\alpha_s({\bar\mu}))\right]\;,
\end{equation}
where $x_1$ and $x_2$ are momentum fractions of partons, $b_1$ and
$b_2$ are conjugate variables of parton transverse momenta $k_{1T}$ and
$k_{2T}$, and $\gamma_\phi$ is the anomalous dimension of mesons. 
The hard part $H(t)$ can be calculated perturbatively.
$C_i(t)$ is the Wilson coefficient corresponding to the four-quark
operator causing $B\to K$ transition. 
The scale $t$ is given explicitly in terms of $x_1$, $x_2$, $b_1$, $b_2$
and $M_B$, and it is of $O(\sqrt{{\bar\Lambda}M_B})$.
Here ${\bar\Lambda}=M_B-m_b$, where $M_B$ and $m_b$ are $B$ meson mass
and $b$ quark mass, respectively.
It is important to note that in PQCD, the scale of the Wilson
coefficient $t$ can reach below $M_B/2$.
In the factorization assumption \cite{BSW}, this scale is fixed at $M_B/2$.
The Wilson coefficient for a penguin operator increases as
the scale evolves down. 
This explains the enhancement of the penguin amplitude in PQCD compared
to the amplitude obtained by the factorization assumption.

\begin{figure}[hbt]
\begin{center}
\includegraphics{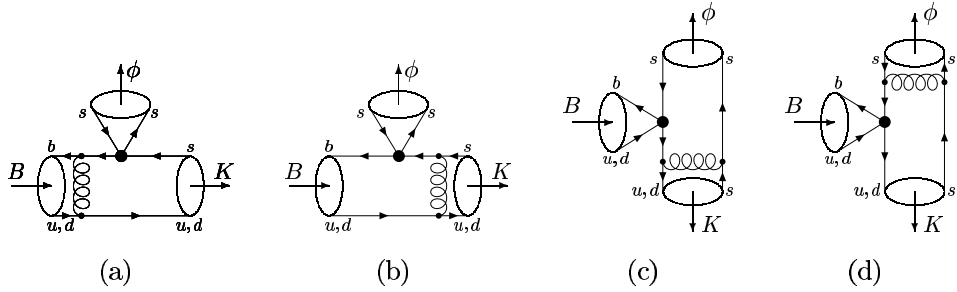}
\end{center}
\caption{Feynman diagrams contributing to factorizable 
amplitudes for $B \to \phi K$}
\label{fig:fact}
\end{figure}

\begin{figure}[hbt]
\begin{center}
\includegraphics{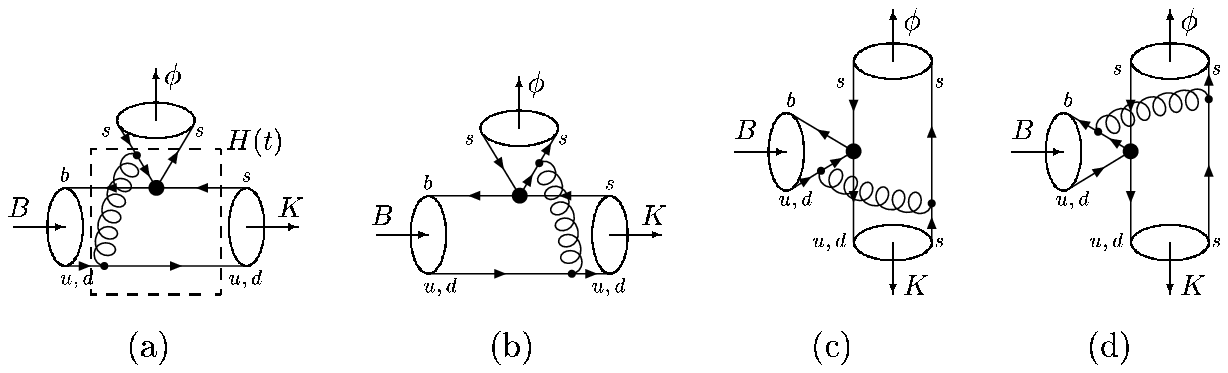}
\end{center}
\caption{Feynman diagrams contributing to nonfactorizable amplitudes for
 $B\to\phi K$}
\label{fig:nonfact}
\end{figure}

In this method, we can calculate not only factorizable amplitudes but
also nonfactorizable and annihilation amplitudes.
In case of $B\to \phi K$ decays, the factorizable amplitudes which can
be written in terms of form factors $F^{BK}$ and $F^{\phi K}$ 
are shown in Fig.~\ref{fig:fact}(a)-(d). The
nonfactorizable amplitudes are shown in Fig.~\ref{fig:nonfact}(a)-(d). 
Ellipses denote meson wave functions in these figures.
For illustration purposes, we show the hard part of the nonfactorizable
diagram as the dashed box in Fig.~\ref{fig:nonfact}(a).
The parameters in meson wave functions are calculated from the
light-cone QCD sum rules, and the theoretical uncertainty of the
parameters is about 30\%.
The hard part depends on the particular processes, but it is calculable.
The wave functions contain non-perturbative dynamics and are not
calculable, but once it is known, it can be used for other decay
processes.  

In this paper, we calculate branching ratios for $B\to \phi K$ modes
using PQCD approach. The detail is discussed in Ref.~\cite{phik}.
We predict the branching ratios for $B\to \phi K$ decays, and our
predictions agree with the current experimental data and are larger than
the values obtained from the naive factorization assumption (FA) and the
QCD-improved factorization (QCDF) \cite{QIF1},~\cite{QIF2}.

%
%

\section{$B\to \phi K$ Amplitudes}\label{sec:amp}
We consider $B$ meson to be at rest.
In the light-cone coordinate, the $B$ meson momentum $P_1$, the $K$
meson momentum $P_2$ and the $\phi$ meson momentum $P_3$ are taken to be
\begin{eqnarray}
P_1=\frac{M_B}{{\sqrt 2}}(1,1,{\bf 0}_T)\;,\;\;
P_2=\frac{M_B}{{\sqrt 2}}(1-r_{\phi}^2,0,{\bf 0}_T)\;,\;\;
P_3=\frac{M_B}{{\sqrt 2}}(r_{\phi}^2,1,{\bf 0}_T)\;,
\end{eqnarray}
where $r_{\phi} = M_{\phi}/M_B$, and the $K$ meson mass is neglected. 
The momentum of the spectator quark in the $B$ meson is written as $k_1$.
Since the hard part is independent of $k_1^+$, 
the $\delta(k_1^+)$ function appears after integrating over its conjugate
spacial variable. Therefore, $k_1$ has only the minus component $k_1^-$
and small transverse components ${\bf k}_{1T}$. 
$k_1^-$ is given as $k_1^-=x_1P_1^-$, where $x_1$ is a momentum fraction.
The quarks in the $K$ meson have plus components $x_2P_2^+$ and
$(1-x_2)P_2^+$, and the small transverse components ${\bf k}_{2T}$ and
$-{\bf k}_{2T}$, respectively. 
The quarks in the $\phi$ meson have the minus components $x_3P_3^-$ and
$(1-x_3)P_3^-$, and the small transverse components ${\bf k}_{3T}$ and
$-{\bf k}_{3T}$, respectively. 
The $\phi$ meson longitudinal polarization vector $\epsilon_{\phi}$ and
two transverse polarization vector $\epsilon_{\phi T}$ are given by 
$\epsilon_{\phi}=(1 / \sqrt{2}r_{\phi})( -r_{\phi}^2, 1, {\bf 0}_T)$
and 
$\epsilon_{\phi T}=(0,0,{\bf 1}_T)$.

The $B$ meson wave function for incoming state and the $K$ and $\phi$
meson wave functions for outgoing state with up to twist-3 terms are
written as  
\begin{eqnarray}
\Phi^{(in)}_{B,\alpha\beta,ij}&=&
\frac{i\delta_{ij}}{\sqrt{2N_c}}\int dx_1 d^2{{\bf k}_{1T}} 
e^{-i(x_1P_1^-z_1^+-{{\bf k}_{1T}}{{\bf z}_{1T}})}
\left[
(\not P_1 + M_B)\gamma_5 \phi_B(x_1,{\bf k}_{1T})
\right]_{\alpha\beta}\;,\\
\Phi^{(out)}_{K,\alpha\beta,ij}
&=&\frac{-i\delta_{ij}}{\sqrt{2N_c}}\int^1_0dx_2 e^{ix_2P_2\cdot z_2}
\gamma_5\left[
\not P_2\phi^A_K(x_2)+m_{0K}\phi_K^P(x_2)
+m_{0K} (\not v\not n -1)\phi_K^T(x_2)
\right]_{\alpha\beta}\;,\\
\Phi^{(out)}_{\phi,\alpha\beta,ij}
&=&\frac{\delta_{ij}}{\sqrt{2N_c}}\int^1_0dx_3 e^{ix_3P_3\cdot z_3}
\left[
M_\phi\not\epsilon_\phi \phi_\phi(x_3)+
\not\epsilon_\phi \not P_3\phi_\phi^t(x_3)
+ M_\phi\phi_\phi^s(x_3)
\right]_{\alpha\beta}\;,
\end{eqnarray}
where $i$ and $j$ is color indices, and $\alpha$ and $\beta$ are Dirac
indices. $m_{0K}$ is related to the chiral symmetry breaking scale,  
$m_{0K}= M_K^2/(m_d+m_s)$. $v$ and $n$ are defined as $v^\mu =
P_2^\mu/P_2^+$ and $n^\mu = z_2^\mu/z_2^-=(0,1,{\bf 0}_T)$.
We neglect the terms which are proportional to the transverse
polarization vector $\epsilon_\phi^T$, because these terms 
drop out from our calculation kinematically.
The explicit form of these wave functions will be shown in
Sec.~\ref{sec:num}.

Widths of $B\to \phi K$ decays can be expressed as
\begin{equation}
\Gamma=\frac{G_F^2}{32\pi M_B}|{\cal A}|^2\;.
\end{equation}
The decay amplitudes, ${\cal A}$, and $\bar{\cal A}$, corresponding to
$B^0\to \phi K^0$, and ${\bar B^0}\to \phi {\bar K^0}$, respectively, are
written as 
\begin{eqnarray}
{\cal A}&=&f_{\phi}V_{ts}V^*_{tb}
F^P_e+V_{ts}V^*_{tb}
{\cal M}^P_e
+f_BV_{ts}V^*_{tb}
F^P_a+V_{ts}V^*_{tb}
{\cal M}^P_a\;,\\
\bar{\cal A}&=&f_{\phi}V^*_{ts}V_{tb}F^P_e+V^*_{ts}V_{tb}{\cal M}^P_e
+f_BV^*_{ts}V_{tb}F^P_a+V^*_{ts}V_{tb}{\cal M}^P_a\;,
\end{eqnarray}
where $F_e$ is the amplitude for factorizable diagrams which are considered
in FA. $F_a$ and ${\cal M}$ are the annihilation factorizable and the
nonfactorizable diagrams which are neglected in FA.  
The indices, $e$, and $a$, denote the tree topology, and annihilation
topology, respectively. 
The index $P$ denotes the contribution from diagrams with a penguin
operator.   
The decay amplitudes, ${\cal A}^+$, and ${\cal A}^-$, corresponding to
$B^+\to \phi K^+$, and $B^-\to \phi K^-$, respectively, are written as
\begin{eqnarray}
{\cal A}^+&=&f_{\phi}V_{ts}V^*_{tb}
F^P_e+V_{ts}V^*_{tb}
{\cal M}^P_e
+f_BV_{ts}V^*_{tb}
F^P_a+V_{ts}V^*_{tb}
{\cal M}^P_a
-f_BV_{us}V^*_{ub}
F^T_a-V_{us}V^*_{ub}
{\cal M}^T_a\;,\\
{\cal A}^-&=&f_{\phi}V^*_{ts}V_{tb}F^P_e+V^*_{ts}V_{tb}{\cal M}^P_e
+f_BV^*_{ts}V_{tb}F^P_a+V^*_{ts}V_{tb}{\cal M}^P_a
-f_BV^*_{us}V_{ub}F_a^T-V^*_{us}V_{ub}{\cal M}^T_a\;,
\end{eqnarray}
where the index $T$ denotes tree contributions. 
Since the Cabibbo-Kobayashi-Maskawa (CKM) matrix elements for the tree
amplitudes are much smaller than for the penguin amplitudes, the tree
contributions are very small. 

The factorizable diagrams are given as Fig.~\ref{fig:fact}. The
factorizable penguin amplitude, $F_e^P$, which comes from 
Fig.~\ref{fig:fact}(a) and Fig.~\ref{fig:fact}(b) is written as
\begin{eqnarray}
F^P_e 
&=&
8\pi C_F M_B^4 \int_0^1 dx_1 dx_2 \int_0^{\infty} b_1db_1 b_2db_2 
\phi_B(x_1,b_1)
\nonumber \\
& & \times
\bigg\{ 
\left[ (1+x_2)\phi_K^A(x_2)+r_K(1-2x_2)
\left( \phi_K^P(x_2)+\phi_K^T(x_2) \right) \right]
E_{e}(t^{(1)}_e) N_t \{x_2(1-x_2)\}^c h_e(x_1,x_2,b_1,b_2)
\nonumber\\
& &\;\;\;\;\;\; +2r_K \phi_K^P(x_2)E_{e}(t^{(2)}_e) 
N_t \{x_1(1-x_1)\}^c
h_e(x_2,x_1,b_2,b_1)
\bigg\}\;,
\end{eqnarray}
where $N_t \{x(1-x)\}^c$ is the factor for the threshold
resummation \cite{TR}. We use $N_t=1.775$ and $c=0.3$ \cite{KuLS}. 
The evolution factors are defined by 
$E_{e}(t)=\alpha_s(t)a_e(t)\exp[-S_B(t)-S_K(t)]$
where $\exp[-S_i(t)]$ is the factor for the $k_T$
resummation \cite{CS},~\cite{BS}. The explicit forms of the factor
$S_i(t)$ are given, for example, in Ref.~\cite{KLS}.   
The hard scales $t^{(1)}_e$ and $t^{(2)}_e$, which are the scales in
hard process, are 
given by  
$t^{(1)}_e={\rm max}(\sqrt{x_2}M_B,1/b_1,1/b_2)$ and $t^{(2)}_e={\rm
max}(\sqrt{x_1}M_B,1/b_1,1/b_2)$. 
The Wilson coefficient is given by
\begin{eqnarray}
a_e(t)=
 C_3+\frac{C_4}{N_c}
+ C_4+\frac{C_3}{N_c} 
+ C_5+\frac{C_6}{N_c} 
-\frac{1}{2}
\left(
C_7+\frac{C_8}{N_c} 
+C_9+\frac{C_{10}}{N_c} 
+C_{10}+\frac{C_9}{N_c}
\right)\;.
\end{eqnarray}
The hard function, which is the Fourier transformation of the virtual
quark propagator and the hard gluon propagator, is given by
\begin{eqnarray}
h_e(x_1,x_2,b_1,b_2)&=&K_{0}\left(\sqrt{x_1x_2}M_Bb_1\right)
\left[\theta(b_1-b_2)K_0\left(\sqrt{x_2}M_B
b_1\right)I_0\left(\sqrt{x_2}M_Bb_2\right)
+(b_1 \leftrightarrow b_2)
\right]\;.
\end{eqnarray}

The factorizable annihilation diagrams shown in Fig.~\ref{fig:fact}(c) and
Fig.~\ref{fig:fact}(d), and the nonfactorizable diagrams shown in
Fig.~\ref{fig:nonfact}(a)-(d) can be also calculated in the same way as
$F_e^P$ \cite{phik}.

%
%

\section{Numerical Results}\label{sec:num}
We use the model of the $B$ meson wave function written as
\begin{eqnarray}
\phi_B(x,b) &=& N_B
x^2 (1-x)^2\exp\left[-\frac{1}{2}
\left(\frac{xM_B}{\omega_{B}}\right)^2
-\frac{\omega_{B}^2 b^2}{2}\right] \;,
\end{eqnarray}
where $\omega_{B}=0.40$ GeV \cite{BW}. $N_B$ is determined
by normalization condition given by 
\begin{eqnarray}
\int_0^1 dx\phi_B(x,b=0)&=&\frac{f_B}{2\sqrt{2N_c}}\;.
\label{eq:bnor}
\end{eqnarray}

The $K$ meson wave functions are given as
\begin{eqnarray}
\phi_{K}^A(x) &=& \frac{f_K}{2\sqrt{2N_c}}6x(1-x)
\left[1 + 3a_1(1-2x)
+ \frac{3}{2}a_2 \left\{ 5(1-2x)^2-1 \right\} 
\right] \;,
\\
\phi^P_{K}(x) &=& \frac{f_K}{2\sqrt{2N_c}}
\bigg[ 
1
+\frac{1}{2}\left(30\eta_3 -\frac{5}{2}\, \rho_K^2\right)
\left\{3(1-2x)^2-1\right\} \nonumber\\
& & -\frac{1}{8}\left( 3\eta_3\omega_3+\frac{27}{20}\, \rho_K^2
  + \frac{81}{10}\, \rho_K^2 a_2\right)
\left\{3-30(1-2x)^2+35(1-2x)^4\right\} 
\bigg]\;,
\\
\phi^T_{K}(x) &=& \frac{f_K}{2\sqrt{2N_c}}(1-2x)
\bigg[ 1
+
6\left(5\eta_3 -\frac{1}{2}\,\eta_3\omega_3 - \frac{7}{20}\,
      \rho_K^2 - \frac{3}{5}\,\rho_K^2 a_2 \right)
(1-10x+10x^2) \bigg]\;,
\end{eqnarray}
where $\rho_K=(m_d+m_s)/M_K$ \cite{PB1},~\cite{PB2}.
The parameters of these wave functions are given as 
$a_1=0.17,\;
a_2=0.20, \;
\eta_3=0.015$ and  
$\omega_3=-3.0$ where the renormalization scale is 1 GeV. 

The $\phi$ meson wave functions are given as
\begin{eqnarray}
\phi_{\phi}(x) &=& \frac{f_{\phi}}{2\sqrt{2N_c}}6x(1-x)\;,
\\
\phi^t_{\phi}(x) &=& \frac{f^T_{\phi}}{2\sqrt{2N_c}}
\bigg[ 3(1-2x)^2
+\frac{35}{4}\zeta_3^T\{3-30(1-2x)^2+35(1-2x)^4\} \nonumber\\
& & +\frac{3}{2}\delta_{+}\left\{1-(1-2x)\log\frac{1-x}{x}\right\} 
\bigg]\;,
\\
\phi^s_{\phi}(x) &=& \frac{f^T_{\phi}}{4\sqrt{2N_c}}
\bigg[ 
(1-2x)\left(
6+9\delta_{+}+140\zeta_3^T-1400\zeta_3^Tx+1400\zeta_3^Tx^2
\right)+3\delta_{+}\log\frac{x}{1-x}
 \bigg]\;,
\end{eqnarray}
where $\zeta_3^T=0.024$ and $\delta_{+}=0.46$ \cite{PB3}. 
We have found that the final results are insensitive to the values
chosen for $\zeta_3^T$ and $\delta_{+}$.

We use the Wolfenstein parameters for the CKM matrix elements   
$A = 0.819, \;
\lambda = 0.2196, \;
R_b \equiv \sqrt{\rho^2+\eta^2}= 0.38$ \cite{PDG}, and choose the angle
$\phi_3 = \pi/2$ \cite{KLS}. 
We have found that the final results are quite insensitive to the values
of $\phi_3$.
For the values of meson masses, we use $M_B = 5.28\; {\rm GeV}$, $M_K =
0.49 \;{\rm GeV}$ and $M_{\phi}=1.02\; {\rm GeV}$. 
In addition, for the values of meson decay constants, we use $f_{B}
= 190\; {\rm MeV}$, $f_{K} = 160\; {\rm MeV}$, $f_{\phi} = 237 \;{\rm
MeV}$ and $f_{\phi}^T = 215 \;{\rm MeV}$. The $B$ meson life times are
given as $\tau_{B^0}=1.55\times 10^{-12}\;{\rm sec}$ and
$\tau_{B^\pm}=1.65\times 10^{-12} \;{\rm sec}$. And we use $\Lambda_{\rm
QCD}^{(4)}=0.250\;{\rm GeV}$ and $m_{0K} =1.70\; {\rm GeV}$ \cite{KLS}. 

We show the numerical results of each amplitude for $B^0 \to \phi K^0$
and $B^\pm \to \phi K^\pm$ decays in Tab.~\ref{table1}. 
The factorizable penguin amplitude $F_e^P$ gives a dominant contribution
to $B \to \phi K$ decays. 
The factorizable annihilation penguin amplitude $F_a^P$ generates a large
strong phase.
In $B^\pm \to \phi K^\pm$ modes, there are  contributions from $f_B
F_{a}^T$ and $M_{a}^T$. These tree amplitudes contribute only a few percent
to the whole amplitude, since the CKM matrix elements related to the
tree amplitudes are very small.
In order to isolate the trivial uncertainty from $f_B$, $f_K$ and
$f_\phi$, we express our prediction for $B \to \phi K$ as
\begin{eqnarray}
{\rm Br}(B^0\to \phi K^0)&=& 
\left| \frac{f_B f_K f_\phi}
{190\; {\rm MeV}\; 160\; {\rm MeV}\; 237\; {\rm MeV}} \right|^2  
\times \left( 9.43\times 10^{-6} \right)\;,   \\
{\rm Br}(B^\pm\to \phi K^\pm)&=& 
\left| \frac{f_B f_K f_\phi}
{190\; {\rm MeV}\; 160\; {\rm MeV} \; 237\; {\rm MeV}} \right|^2
\times \left( 10.1\times 10^{-6} \right)\;.
\end{eqnarray}
We found that our result is insensitive to $f_\phi^T/f_\phi$. For
example, 10\% variation of $f_\phi^T/f_\phi$ leads to less than 1\%
variation in our final result. 
The current experimental values are summarized in Tab.~\ref{table2}. 
The values which are predicted in PQCD are consistent with the current
experimental data. 
However, our branching ratios have the theoretical error from the
$O(\alpha_s^2)$ corrections, the higher twist corrections, and the error
of input parameters. Large uncertainties come from the meson decay
constants, the shape parameter $\omega_B$, and $m_{0K}$. 
These parameters are fixed 
from the other modes ($B\to K\pi,\; D\pi$, etc.).   
We try to vary $\omega_B$ from 0.36 GeV to 0.44 GeV, then we obtain 
Br($B^\pm\to \phi K^\pm) = (7.54 \sim 13.9)\times 10^{-6}$. 
Next, we set $\omega_B=0.40$ and try to vary
$m_{0K}$ from 1.40 GeV to 1.80 GeV, then we obtain  Br($B^\pm
\to \phi K^\pm) = (6.65 \sim 11.4)\times 10^{-6}$.

Now, we consider the ratio of the branching ratio for the $B^0 \to \phi K^0$
decay to the one for the $B^+ \to \phi K^+$ decay. 
The theoretical uncertainties from various parameters are small, since the
parameters in the numerator cancel out those in the denominator.
The difference between the two branching ratios come in principle from
$B$ meson 
life times, tree and electroweak penguin contributions in
annihilation amplitudes. 
We found that the tree and electroweak penguin amplitudes in the
annihilation diagrams are negligible.
Tree amplitudes are suppressed by two factors. 
First, they are annihilation processes which are suppressed by helicity.  
Second, they are multiplied by small CKM matrix elements.
We predict that this ratio is
\begin{eqnarray}
\frac{{\rm Br}(B^0 \to \phi K^0)}{{\rm Br}(B^+ \to \phi K^+)}=0.95\;,
\end{eqnarray}
where the theoretical uncertainties from $m_{0K}$ and
$\omega_B$ are less than $1\%$.    
The ratio is essentially given by the life time difference.
The experimental value of this ratio from BELLE \cite{BELLE} 
is ${\rm Br}(B^0 \to \phi K^0)/{\rm Br}(B^+ \to \phi K^+)
 = 0.82^{+0.39}_{-0.32}\pm 0.10$.  

In FA, the branching ratio is very sensitive
to the effective number of colors $N_c^{eff}$. If we set $N_c^{eff}$=3,
then the branching ratio is about $4.5\times 10^{-6}$ where the scale of
the Wilson coefficient is taken to $M_B/2$ and $F^{BK}$ is 0.38 from the
BSW model. 
In QCDF, branching ratios for $B \to \phi K$ decays are predicted as 
Br$(B^0 \to \phi K^0) = (4.0^{+2.9}_{-1.4})\times 10^{-6}$ and 
Br$(B^- \to \phi K^-) = (4.3^{+3.0}_{-1.4})\times 10^{-6}$ 
including the annihilation diagram \cite{QIF2}. 
Our predicted values are larger than these results.
This is due to the enhancement of the Wilson coefficient for the penguin 
amplitude as explained in Sec.~\ref{sec:intro}.
In PQCD approach, the scale of the Wilson coefficients, which is equal
to the hard scale $t$, can reach lower values than $M_B/2$.

%
%

\section{Summary}
In this paper, we calculate $B^0\to \phi K^0$ and $B^\pm\to \phi K^\pm$
decays in PQCD approach. Our predicted branching ratios agree 
with the current experimental data and are larger than the values
obtained by FA and QCDF. Because the Wilson coefficients for penguin
operators are enhanced dynamically in PQCD.


\vspace{0.4cm}
\noindent{\bf Note added:}
           
After this work has been completed, we become aware of a similar
calculation by Chen, Keum and Li \cite{CKL}.
Our results are in agreement.


\vspace{0.4cm}
\centerline{\bf Acknowledgment}
\vspace{0.5cm}

The topic of this research was suggested by Professor A.I.~Sanda. 
The author thanks his PQCD group members: Y.Y.~Keum, E.~Kou, T.~Kurimoto,
H-n.~Li, T.~Morozumi, R.~Sinha, K.~Ukai for useful discussions.
The author thanks JSPS for partial support.
The work was supported in part by Grant-in Aid for Special Project
Research (Physics of CP violation); Grant-in Aid for Scientific Research
from the Ministry of Education, Science and Culture of Japan.




\begin{table}[hbt]
\begin{center}
\begin{tabular}{ccc}
\hline
&  $B^0 \to \phi K^0$ & $B^\pm \to \phi K^\pm$ 
\\
\hline
$f_\phi F_e^P$ & 
 $-1.03 \times 10^{-1}$ & 
 $-1.03 \times 10^{-1}$ 
\\
$f_B F_{a}^P$ & 
 $ 6.45 \times 10^{-3} + \,\, i \,\, 4.28 \times 10^{-2}$ &
 $ 6.17 \times 10^{-3} + \,\, i \,\, 4.20 \times 10^{-2}$ 
\\
$M_{e}^P$ & 
 $ 5.24 \times 10^{-3} - \,\, i \,\, 3.61 \times 10^{-3}$ &
 $ 5.24 \times 10^{-3} - \,\, i \,\, 3.61 \times 10^{-3}$ 
\\
$M_{a}^P$ & 
 $-8.03 \times 10^{-4} - \,\, i \,\, 1.73 \times 10^{-3}$ &      
 $-6.56 \times 10^{-4} - \,\, i \,\, 7.22 \times 10^{-4}$ 
\\
$f_B F_{a}^T$ & 
 &
 $-1.11 \times 10^{-1} - \,\, i \,\, 3.75 \times 10^{-2}$ 
\\
$M_{a}^T$ & 
 &
 $ 1.60 \times 10^{-2} + \,\, i \,\, 2.77 \times 10^{-2}$ 
\\
\hline
\end{tabular}
\end{center}
\caption{Contribution to $B^0 \to \phi K^0$ and 
$B^\pm \to \phi K^\pm$ decays from each amplitude}   
\label{table1}
\end{table}

\begin{table}[hbt]
\begin{center}
\begin{tabular}{ccc}
\hline
& ${\rm Br}(B^0\to \phi K^0)$ & ${\rm Br}(B^\pm\to \phi K^\pm)$
\\
\hline
BaBar &
$(8.1^{+3.1}_{-2.5}\pm 0.8)\times 10^{-6}$ &
$(7.7^{+1.6}_{-1.4}\pm 0.8)\times 10^{-6}$
\\
BELLE &
$(8.7^{+3.8}_{-3.0}\pm1.5)\times 10^{-6}$ &
$(10.6^{+2.1}_{-1.9}\pm2.2)\times10^{-6}$
\\
CLEO &
$ < 12.3 \times 10^{-6}$ &
$(5.5^{+2.1}_{-1.8}\pm 0.6)\times 10^{-6}$
\\
\hline
\end{tabular}
\end{center}
\caption{The experimental data of $B \to \phi K$ branching ratios
 from BaBar\protect\cite{BABAR}, BELLE\protect\cite{BELLE} and
 CLEO\protect\cite{CLEO}}    
\label{table2}
\end{table}

\end{document}